# Photo-degradation Protection in 2D In-Plane Heterostructures Revealed by Hyperspectral Nanoimaging: the Role of Nano-Interface 2D Alloys

**Alireza Fali[1], Tianyi Zhang[2], Jason Patrick Terry[1], Ethan Kahn[2], Kazunori Fujisawa[3], Sandhaya Koirala[1], Yassamin Ghafouri[1], Wenshen Song[4], Li Yang[4], Mauricio Terrones[2,3,5], Yohannes Abate[1*]**

[1] Department of Physics and Astronomy, University of Georgia, Athens, Georgia 30602
[2] Department of Materials Science and Engineering, The Pennsylvania State University, University Park, PA 16802
[3] Department of Physics, The Pennsylvania State University, University Park, PA 16802
[4] Department of Physics and Institute of Materials Science & Engineering, Washington University in St Louis, St. Louis, MO 63130
[5] Department of Chemistry, The Pennsylvania State University, University Park, PA 16802

* Corresponding author: Mauricio Terrones (mut11@psu.edu), Yohannes Abate (*yohannes.abate@uga.edu*)

## Abstract
Single-layer heterostructures exhibit striking quasiparticle properties and many-body interaction effects that hold promise for a range of applications. However, their properties can be altered by intrinsic and extrinsic defects, thus diminishing their applicability. Therefore, it is of paramount importance to identify defects and understand 2D materials' degradation over time using advanced multimodal imaging techniques as well as stabilize degradation via built-in interface protection. Here we implemented a liquid-phase precursor approach to synthesize 2D in-plane $MoS_2$-$WS_2$ heterostructures exhibiting nanoscale alloyed interfaces and map exotic interface effects during photo-degradation using a novel combination of hyperspectral tip-enhanced photoluminescence, Raman and near-field nanoscopy. Surprisingly, 2D alloyed regions exhibit remarkable thermal and photo-degradation stability providing protection against oxidation. Coupled with surface and interface strain, 2D alloy regions create localized potential wells that concentrate excitonic species via a charge carrier funneling effect. These results provide a clear understanding of the importance of 2D alloys as systems able to withstand degradation effects over time, and could be now used to stabilize optoelectronic devices based on 2D materials.

## MAIN TEXT

### Introduction

The creation of in-plane and out-of-plane van der Waals (vdW) heterostructures from atomically thin monolayers has opened unprecedented possibilities to artificially design excitonic materials with strong many-body effects that exhibit exciting new physics and novel applications.(*1, 2*) Building vdW heterostructures is like assembling an atomic Lego® set where each two-dimensional (2D) crystal Lego block can be vertically stacked, rotated or laterally stitched together to produce structures with designed excitonic effects. (*2, 3*) Both top-down and bottom-up strategies have been used to assemble complex excitonic vdW heterostructures with new properties not found in the bare monolayers. (*3-5*)

In-plane heterostructures with *lateral* interfaces, distinct from vertical heterostructures, can be created via covalent bonds of different domains as a single, continuous sheet that can then be extended into vdW stacks of more complex heterostructures. These heterojunctions could result in fascinating physical phenomena for constructing unique devices based on high-mobility field-effect transistors and planar monolayer p–n junctions. (*6*) In these lateral interfaces, new types of interactions such as intralayer coupling, lateral strain, interface defects and 2D alloys can have significant effects on the oxidation process and spatial distribution of exciton emission. The heterojunction boundary between dissimilar lateral domains presents a novel platform to explore interface physics in one dimension such as spin and valley transport and exciton diffusion across the interface, many-body interaction effects in 1D quantum confined quasiparticles trapped between atomically thin barriers. (*7*) In vertically stacked heterostructures, photo-induced oxidation and environmental effects on exciton species are inherently confined to a single layer or interlayer of the heterostructure. Due to the advantages these heterointerfaces offer, new approaches to monitor and understand their stability and degradation over time are needed, and multimodal advanced imaging and spectroscopy techniques with ultrahigh resolution could be implemented. These techniques are advantageous due to their sensitivity and ability to initiate and probe light-driven nanoscale optical and electronic processes in 2D systems. Consequently, high sensitivity photoluminescence (PL) /Raman spectroscopy and imaging techniques have been applied to study a wide range of exciton related phenomena in 2D materials and their heterostructures.(*8-15*) However, it is still important to perform a comprehensive understanding of the surface and interface chemistry of 2D heterointerfaces, as well as their mechanical, electronic and optical effects that give rise to the complex excitonic phenomena and monitor their evolution in time with high spatial resolution.

In this work, using in-plane $MoS_2$-$WS_2$ heterostructures (with Mo-rich in inner regions, W-rich in outer regions, and a ~100 nm 2D alloyed interface. See details in Fig. 1) as a model system, we observed interface-protected exciton stability, localization and inhomogeneities during photodegradation by implementing a combination of hyperspectral tip-enhanced photoluminescence and Raman nanoimaging in concert with scattering type scanning near-field microscopy. Monosulfur vacancies abundant along the edge regions offer thermodynamically favorable sites for oxygen reactivity, favoring the photo-induced oxidation of the outer $WS_2$ monolayer (ML) while oxidation progression to the inner $MoS_2$ ML is inhibited by the presence of the 2D alloy interface. (*16*) Interestingly, this alloy interface corner regions, being thermodynamically stable (*17*) , exhibit higher stability against photooxidation and facilitate the generation of localized potential wells that accumulate excitonic species via charge carrier funneling effect with minimized Auger recombination losses. The spectral weight of trion and biexciton, when compared to the neutral exciton intensity, decreases with photo-degradation, and biexciton emission quenches much faster than trion emission due to changes in Coulomb screening caused by the oxidation process. High-resolution scanning transmission electron microscopy (HRSTEM) characterizations revealed that $WS_2$-$MoS_2$ alloy regions are stable during and after exposure at high temperatures. These results provide novel insights into the role of the 2D alloys at lateral interfaces and evolution of neutral and charged excitons critical for optical, electrical and thermal device engineering based on 2D materials and their heterostructures.

## Results

The samples used for this work are in-plane $MoS_2$-$WS_2$ heterostructures grown on $SiO_2$/Si by spin-coating of liquid-phase precursors and subsequent high-temperature sulfurization under ambient pressure (see Methods for the detailed process). (*14*) Optical microscopy (Fig. 1A) shows the monolayer exhibiting a truncated triangular shape. The optical image reveals that the center and edge regions exhibit distinct optical contrasts, separated by an interface region highlighted by the white dashed line, which indicates different chemical compositions within center and edge regions of the monolayer. Far-field Raman and PL spectroscopies were used to characterize structural and optical properties of the $MoS_2$-$WS_2$ heterostructures, respectively. PL spectra (Fig. 1B) in the hetero-monolayer reveal PL emission at ~1.81 eV and ~ 1.94 eV from center and edge, close to the optical band gap of $MoS_2$ and $WS_2$ monolayers, respectively. (*18*) Raman spectroscopy (Fig. 1C) reveals spectroscopic features of $MoS_2$ E' and $A_1$' modes at the center, and $WS_2$ E' and $A_1$' modes at the edge while the interface region displays a combination of $MoS_2$-like and $WS_2$-like Raman modes. Aberration corrected high-resolution scanning transmission electron microscopy (*AC-HRSTEM*) images in Fig.1D-F provide atomic-resolution images of the $MoS_2$-$WS_2$ heterostructures; Mo (indicated by red circles) and W atoms (indicated by green circles) can be clearly resolved by the atomic number contrast in the high-angle annular dark-field (HAADF) imaging mode. AC-HRSTEM confirms that the center and edge regions are Mo-rich and W-rich, respectively, and also reveals some W alloying in the central Mo-rich region (Fig. 1D&F). It should be noted that the STEM image at the interface between Mo-rich and W-rich regions unambiguously shows the intermixing of Mo and W atoms, forming a region of alloyed $Mo_xW_{1-x}S_2$ (Fig. 1E). The width of the alloyed interface is on the order of ~100 nm, according to our recent observations on the $MoS_2$-$WS_2$ heterostructures grown using identical conditions.(*19*)

The schematics of the multi-modal imaging setup are shown in Fig. 1G. The integrated near-field, hyperspectral TEPL and TERS nano-imaging technique we introduce here is a combination of a commercial s-SNOM (neaspec co.) with a grating spectrometer, coupled to a silicon charge-coupled device (iDUS, Andor). This configuration allows detecting either the elastic backscattered field, as typically of s-SNOM, or inelastic forward scattered signals. The tip-enhanced inelastic forward scattered signals (TEPL or TERS) are dispersed in a spectrometer, after passing through a notch filter to remove elastically scattered field, and detected with a CCD (see Methods for details).

We demonstrate tip-enhanced hyperspectral PL nanoimaging of ML in-plane heterostructure and map the temporal and spatial degradation and exciton evolution over an extended time (> 100 days). Figure 2 shows 3D hyperspectral data cube taken at 1, 5, 20, 85 and 103 days since the start of experiment acquired by measuring an array of 85 by 85 pixels TEPL normalized spectra of the ML heterostructure (shown in Fig. 1A). The x and y axes of the 3D data cube shown Fig. 2A, indicate the plane of the sample surface while the z-axis correspond to the photon-energy axis (1.77 eV to 2 eV). The acquisition time for each spectrum was 1 second, thus yielding a total acquisition time of 2 hours per image in Fig. 2A (see Methods for details). Monochromatic TEPL images are then extracted by cutting the cube at selected energies, 1.81 eV and 1.94 eV, close to PL resonance emission of neutral excitons of ML $WS_2$ (Fig. 2B) and $MoS_2$ (Fig. 2C), respectively. These hyperspectral time-series images reveal dissimilar spatio-spectral inhomogeneities between the outer $WS_2$ and inner $MoS_2$ MLs, providing rich information on the evolution of exciton emissions and degradation of the heterostructure. In Figs. 2A and B we observe striking spatially inhomogeneous degradation of the outer $WS_2$ ML neutral exciton emission which evolves over time and completely disappears after several days. This contrasts the evolution of the inner $MoS_2$ ML that shows a homogeneous quasi-stable exciton emission (Fig. 2C). The TEPL image of $MoS_2$ on day 1 also shows signal on $WS_2$ region due the broad and strong $WS_2$ spectrum, when the $WS_2$ peak is subtracted emission only from $MoS_2$ ML is observed (see fig. S1).

The TEPL images in Fig. 2 clearly show that quenching of excitonic emission preferentially begins from the outer $WS_2$ ML and propagates towards the inner $MoS_2$ ML. This may be due to the presence of higher S vacancy concentrations along the edges in CVD grown samples, which lead to accelerated degradation. Our recent atomic-resolution imaging revealed that the areal density of monosulfur vacancies in $WS_2$ is much larger near the edges (~0.92 $nm^{-2}$) than in the interior (~0.33 $nm^{-2}$). (*16*) The large presence of S vacancies along edges presents a thermodynamically favorable energy landscape for oxygen substitutions, greatly accelerating the photo-induced oxidation of the outer $WS_2$ ML (Fig. 2B). Surprisingly, the inner $MoS_2$ ML has shown more resilience to oxidation, exhibiting only ~8% quenching of PL emission on day 103 (compared to Day 1 in Fig. 2B), while the PL of outer $WS_2$ completely disappears during the same period and the same laser exposure. This is interesting because the inner $MoS_2$ region is also expected to have a large number of S vacancies, but the oxidation, unlike $WS_2$, is uniform and minimal. (*20*) Li et al. have also reported a laser-induced oxidation process in TMDC crystals that starts from the edges leaving central area intact, an effect attributed to highly defective edges. (*21*) We note in our work that there is a 2D alloy interface that plays a critical role in making the material it encloses more stable, prohibiting oxidation propagation from the edge inward to $MoS_2$ (see Fig. 5 discussion below). In Fig. 2D, we show high-resolution TEPL hyperspectral 3D data cube of a small section of the heterostructure sample (blue dashed area indicated in Topography Fig. 2D) and identify the interface region between $MoS_2$ and $WS_2$ MLs (Fig. 2D). The alloy interface is characterized by a broad and slightly red shifted spectrum compared to $WS_2$ owing to its alloy nature (fig. S2). By subtracting the $WS_2$ and $MoS_2$ resonance PL emissions from the interface emission we can extract the nano-interface 2D PL image that exists at the physical boundary of $MoS_2$ and $WS_2$ (Fig. 2D) (also see fig. S2). Furthermore, in Fig. 1E, AC-HRSTEM image at the nanoscale (~100 nm) alloyed interface shows an intermixing of $MoS_2$ and $WS_2$ (local concentrations of ~35% W and ~65% Mo). The formation of $Mo_xW_{1-x}S_2$ alloys is energetically favored and thus thermodynamically stable (*17*), which may result in higher stability against photooxidation, thus indicating a potential route to degradation prevention by lateral 2D encapsulation.

The observed distinctly different ambient degradation behavior within different regions of the ML heterostructure is selectively photo-induced. This is clearly evident in Fig. 3A,B, where light emission of identical flakes that are either exposed to tip-focused laser radiation or ones that are not exposed to laser radiation are compared. In the optical image in Fig. 3A, the laser-irradiated flake is distinguished and marked with a yellow dashed circle, and exhibits a different optical contrast compared with other unexposed flakes. In Fig, 3B the corresponding diffraction limited PL image acquired using a band-pass filter centered at 630 nm, close to the neutral exciton emission of monolayer $WS_2$, clearly shows that the irradiated flake's PL emission from the edge region is completely quenched while other flakes show no signs of oxidation during the span of the experiment. Ambient photooxidation of TMDC has been extensively investigated by several groups (*21-26*) and evidently such selective photooxidation indicate a possible means to use tip nano-focused light for local exciton engineering.

The photooxidation process also induces significant and inhomogeneous topographic and permittivity changes in the ML heterostructure. The topography scan at the start of experiment shows a thickness within ~ 1 nm (Fig. 3C), indicating the monolayer nature of the heterostructures. After photooxidation the topography of the edge $WS_2$ ML increased to 3 nm whereas the inner $MoS_2$ heterostructure increased to ~1.2 nm (Fig. 3E), as shown in the line profile in Fig. 3G. Over a factor of two thickness increase of $WS_2$ ML compared to inner $MoS_2$ ML indicates a stronger photo-induced aging of the $WS_2$ ML region. (*22, 27*) Such topographic height increase has been reported previously and is attributed to laser-induced formation of few nanometer tungsten oxide, as confirmed by Raman measurements performed in the ambient conditions. (*21*) Other studies have attributed topographic changes in photo-oxidized $WS_2$, likely due to the $H_2O$ moisture intercalation effect (*28*), or the alteration of chemical composition in this region. (*22*). In our work, the mechanism of degradation is due to the in-plane heterostructure nature of the ML and the significant role the interface (2D alloy) plays. The alloy interface, located between the outer $WS_2$ and the inner $MoS_2$ regions, protects $MoS_2$ from faster degradation resulting in the outer $WS_2$ ML disproportionate topographic increase.

Simultaneously with topographic images, we acquired near-field s-SNOM amplitude images at the fourth harmonic ($A_4$) of the tip oscillation frequency. Near-field amplitude images taken at day 1 before photooxidation (Fig. 3D) and on day 103 images after photooxidation (Fig. 3F) show striking contrasts. On day 1, the center $MoS_2$ ML show slightly brighter contrast compared to the edge $WS_2$ ML owing to the larger permittivity of $MoS_2$ at 532 nm excitation wavelength compared to $WS_2$ (Fig. 3D, H). (*29*) On day 103, the center $MoS_2$ ML shows significantly brighter contrast compared to the edge $WS_2$ ML (Fig. 3F, H). This is because photooxidation significantly lowers the dielectric constant of $WS_2$ while the inner $MoS_2$ ML remains stable since it is protected by the alloy interface. We analyzed the s-SNOM contrast further by performing calculations using the well-established finite dipole model.(*30*) On day 1 the experimental *4th* harmonic amplitude ratio of the ML to $SiO_2$ substrate are, $A_4(MoS_2)/A_4(SiO_2)$~1.77 and $A_4(WS_2)/A_4(SiO_2)$~1.65. The amplitude contrast calculations using the finite dipole model employing literature data (*29*) for the dielectric values of $MoS_2$ ($\varepsilon = 20 + 11i$) and $WS_2$ ($\varepsilon = 18 + 7i$) and $\varepsilon = 2.15$ for $SiO_2$ gives $A_4(MoS_2)/A_4(SiO_2)$~1.76 and $A_4(MoS_2)/A_4(SiO_2)$~1.68 in good agreement with experiment. On day 103, after photooxidation has occurred, the experimental *4th* harmonic amplitude ratio are $A_4(MoS_2)/A_4(SiO_2)$~1.70 and $A_4(WS_2)/A_4(SiO_2)$~1.27. To match this experimental amplitude ratio, in the calculation we adjusted the dielectric constant for $WS_2$ to $\varepsilon = 6.3+2.5i$, a large decrease from day 1 values used. Thus, the change in the amplitude contrast in $WS_2$ can be explained by its large decrease in the dielectric values due to photooxidation whereas the amplitude contrast for $MoS_2$ remain relatively the same and so is its dielectric values. The near-field amplitude contrast changes indicate that the outer $WS_2$ monolayer has significantly undergone photo-induced oxidation when compared to the inner $MoS_2$ ML. These observed large differences in both optical constants and topographic protrusions between the two MLs as revealed by s-SNOM, despite similar laser exposure and ambient conditions, highlight the significant role of the interface.

To correlate TEPL images (Fig. 2) that show site selective enhanced intensities with structural changes that occur during photooxidation, we performed tip-enhanced Raman (TERS) hyperspectral imaging and acquired spatial and spectral information at every pixel on the same scan area and time sequence as in Fig. 2. In Fig. 4 we show the 3D TERS hyperspectral data cube and monochromatic TERS images taken at 3 selected dates, day 20, 85 and 103, acquired by measuring a TERS spectra array of 60 by 60 pixels. Data acquisition and spectrum normalization were performed similarly to TEPL data cube (see Methods). The gap in the 3D data cube in Fig. 4A is deliberately made to show both the characteristic Raman modes $A'_1$ (very top slice) and E' (top of the cube) of $MoS_2$. For all the selected days shown, we observe intensity changes over time and spatial variations of the E' Raman mode on the outer $WS_2$ and both the E' and $A'_1$ modes on the inner $MoS_2$ that are in a striking similarity with the TEPL images shown in Fig. 2. These similarities indicate that areas containing higher concentration of $WS_2$ or $MoS_2$ inferred from TERS images also show higher PL intensities. (*31*) Despite the same laser exposure, while all typical Raman peaks of $WS_2$ disappear after day 85 (Fig. 4), the intensity of $A'_1$ and E' modes of $MoS_2$ still show robust stability, indicating the influence of the interface (as discussed above) resulting in distinct degradation evolution of $WS_2$ and $MoS_2$ in-plane heterostructures. We have also extracted from the TERS hyperspectral 3D data (Fig. 4A) Raman images at 707 and at 328 $cm^{-1}$ (fig. S3) which are characteristics of monoclinic (O-W-O stretching modes), indicating local oxidation.

Regarding the photodegradation evolution of the 2D heterointerface, figs. 5A-B show PL point spectra of $WS_2$ ML taken at points 1&2 (black dots in Fig. 5C) on different days. The $WS_2$ PL peak at Point 1 dramatically blue shifts by as much as ~ 10 nm after day 5 (red curve) compared to the peak position at day 1 (brown curve). After day 85, the edge peak at Point 1 completely disappears. A similar trend of blue shift, as well as disappearance of the edge $WS_2$ peak, is also observed at Point 2 (weighted average peak position shifts as a function of time are also shown in SI fig. S4). These shifts of exciton energies can be attributed to the degradation-induced change of strain conditions in samples. Previous first-principles many-body perturbation theory (MBPT) studies have shown that the exciton energy is sensitive to strain conditions in monolayer TMDs. (*32*) Thus, we fit the observed shifts of PL peaks of the inner and outer parts in Fig. 5 to those calculated exciton energies of strained monolayer $MoS_2$ and $WS_2$, respectively. (*32*) Figures 5C&D present the fitted strain-distribution maps from the neutral exciton emission on of both Day 1 and Day 20 of the heterostructure sample, respectively. In Fig. 5C, significant strain is observed because of lattice mismatch and a built-in strain in CVD-grown TMDs arising from thermal expansion coefficient mismatch between TMD and substrate. This is the reason why transfer of samples to another substrate has shown to release this built-in strain. (*33*) There are large positive strains (dark red) around the outer $WS_2$ area and very small strain (white) around the inner $MoS_2$ area. For example, the strain is +0.12% at the outer point while that is nearly zero at the inner point, as shown in Fig. 5C. This inhomogeneous strain may be because the outer $WS_2$ structure has a free boundary, making it easier to be strained by the lattice mismatch than the inner part with a connecting alloyed interface and a nearly "fixed" boundary condition. After photooxidation, the outer strain is significantly released. As shown in Fig. 5D the strain at the outer point is reduced to be around 0.02 % on Day 20, which is well within the simulation error bars. This strain release may be from degrading processes. The degradation is more significant in the $WS_2$ part because of its exposed, vulnerable edges and large structure distortions. Where as the change in strain is absent and so is oxidation in the $MoS_2$ part due to protection by the alloyed interface.

We observe from AC-HRSTEM images, alloy interfaces at $WS_2$ crystal corner regions that are close to sharp corners of the inner monolayer $MoS_2$ crystal (Fig. 1E). These alloyed corners and interfaces lead to novel effects. First the corner regions show enhanced stability when compared to the side regions of $WS_2$, as shown in TEPL images on days 5 and 20. These results indicate that the photooxidation of the outer monolayer $WS_2$ crystals is site selective, and avoids regions close to sharp corners of the inner monolayer $MoS_2$ crystals. A similar enhanced stability at corner regions are also shown in the TERS image (Fig. 4B), where the intensity of E' mode of ML $WS_2$ shows preferential higher intensity at corner regions on day 20. A higher intermixing of $MoS_2$ and $WS_2$ into alloys at corner regions is energetically favored and thus thermodynamically stable.

Furthermore, corner regions with enhanced alloy concentration also lead to more local strain due to the atomic size mismatches (atomic radius or bond length). It has been shown that coupling of Mo/W with defects can induce lattice distortions that increase strain. (*34*) We note that, as shown in Fig. 5A&B, point 1 starts at lower energy on day 1, since the corner regions are at higher strain when compared to side regions in $WS_2$ before photooxidation (fig. S5). As the sample degrades, $WS_2$ outer strain is significantly released, but the corner regions remain with higher strain than the rest. This results in concentrating excitons or charge carriers, as revealed in PL inhomogeneous spatial distribution observed in our experiments. Engineering the band structure of two-dimensional materials by external strain has been demonstrated; (*35-39*) the origin and modulation of strain in our case however is intrinsic and dynamic due to the role of the alloy interface and the degradation process. The high-strain corner regions act as potential wells that create exciton reservoirs due to electron and exciton funneling effects (see Fig. 5E) (*40, 41*), allowing localized selective stability of these corner regions, as observed in Fig. 2 hyperspectral TEPL images.

In order to fit and analyze the TEPL spectra of the heterostructure at a corner point (1) and a side point (2) shown in Fig. 5. All three excitonic species $X^0$, $X^-$, and XX can be tracked all the way to cryogenic temperature at both of these selected points during the photooxidation process, as shown in the deconvoluted spectral fitting in fig. S6. The relative PL spectral weights of the species shown in Table S1 reveal a remarkable concentration of trions and biexcitons at day 1 at the corner potential well region (1) compared to the side region (2) on $WS_2$. As the sample degrades however, the trion and biexciton concentration decreases, neutral excitons dominate the emission and the spectral weight of trions to neutral excitons ($X^-/X^0$) decreases, followed by quenching of all excitonic emission as the sample completely oxidizes. This agrees with trends observed by Tongay et al. and others (*42, 43*) , and suggests an increased defect density. It is worth noting that $X^-/X^0$ decreased from 1.8 on day 1 to 0.51 on day 5, whereas $XX/X^0$ decreased much faster from 0.98 to 0.07 in the same time period (Table S1). Because the distance between the two excitons that make up a biexciton is large (3-4 nm), Coulomb screening can easily be modified by oxidation, so that the spectral weight ratio $XX/X^0$ can serve as a sensitive gauge of the photooxidation. (*44*)

The effect of the lateral interface we described above was investigated by limiting the excitation laser power to ~ 400 μW in order to avoid laser-burned holes or similar physical damage to the heterostructure.(*21*) We further extended our investigation and studied the stability of the $Mo_xW_{1-x}S_2$ alloy at elevated temperature by performing repeated HRSTEM scans at different temperatures and alloy compositions. Under this condition, as the continuous electron beam raster scans the surface, a defect could be initiated as a sulfur vacancy due to the combined effect of electronic excitation and knock-on damage, and then the accumulation of electron beam-generated defects leads to the formation of large holes (*45*) . Three regions with different alloying composition were scanned by the electron beam at 400 $^oC$, and the STEM images shown in Fig. 6A reveal $Mo_xW_{1-x}S_2$ alloys just before defect formation and after 126 s. Regardless of the local alloying composition, large holes are generated after several scans. However, for the W-rich region with low intermixing degree ($W_{0.99}Mo_{0.01}S_2$), holes expanded faster than the other two cases with higher intermixing degree. This observation also indicates the higher thermal stability of the 2D alloy. The *in situ* stability test was also carried out at varied temperatures and temporal changes of the hole area is plotted Fig.6B. Although some holes expand faster, holes in regions with higher alloying degree expands slower, especially at low temperatures (400-500 $^oC$). These results indicate the enormous potential of 2D alloy interfaces and alloyed edges for degradation protection of 2D materials.

## Conclusion

Nanoscale alloyed interfaces present a unique platform to stabilize 2D materials over an extended time. We implement integrated TEPL, TERS and s-SNOM techniques based on tapping mode AFM and perform high-resolution hyperspectral imaging of $MoS_2$-$WS_2$ in-plane ML heterostructures exhibiting nanoscale alloy interfaces. The ambient oxidation process, accelerated by the probe laser, begins from the outer ML and is mainly induced by the S vacancies in $WS_2$. The evolution of the degradation is however hampered by the alloy nano-interface region from propagating towards the inner $MoS_2$ ML of the sample. Eventually, the outer ML completely oxidizes, changing from $WS_2$ into W-based oxides while the inner $MoS_2$ exhibits extended stability due to protection by the interface. *In-situ* HRSTEM imaging showed that alloying leads to high stability of 2D materials even at elevated temperatures. The oxidation process causes significant non-uniform changes in the topography and permittivity of the heterostructure. These processes not only cause quenching of neutral and charged exciton emissions, but also induce complex interactions among them that are manifested in neutral exciton, trion and biexciton energy shifts. Notably, the 2D alloy interface coupled with intrinsic strain cause spatial inhomogeneity of the oxidation and emission of the various excitonic species, providing localized potential wells at corner interfaces for various charge carriers and enabling localized emission with enhanced stability. We envision our studies offer a deeper understanding of 2D alloys as a rich platform to engineer stable optoelectronic devices based on 2D heterostructures with enhanced performance.

**Methods:**

**The growth of in-plane $MoS_2$-$WS_2$ heterostructures:** To synthesize in-plane $MoS_2$-$WS_2$ heterostructures, powders of ammonium metatungstate hydrate ($(NH_4)_6H_2W_{12}O_{40}$, 50 mg), ammonium heptamolybdate ($(NH_4)_6Mo_7O_{24}$, 5 mg) and sodium cholate hydrate ($C_{24}H_{39}NaO_5 \cdot xH_2O$, 200 mg) were dissolved in 10 mL of deionized water. The solution was then spin-coated onto a clean $SiO_2$/Si wafer, followed by a single-step high-temperature heating treatment in the chemical vapor deposition (CVD) furnace in the presence of sulfur vapor. During the heating process, the furnace was first ramped to 700 °C and held for 10 min, and subsequently ramped to 800 °C and held for another 10 min. An argon flow of 100 sccm was used as the carrier gas during synthesis.

**Integrated s-SNOM/TEPL/TERS nanoscope:** In the multimodal nanoscope experimental setup (neaspec co.) the sample is excited by a collimated cw green laser (532 nm wavelength). A parabolic mirror with an NA of 0.4 focuses the excitation laser to a commercial PtIr-coated cantilevered Si tips with a vertical oscillation frequency of 240 kHz and amplitude of ~20 nm. Standard near-field retraction curve and near-field nano-imaging is performed by detecting the backscattered near-field via a combination of phase interferometric detection and demodulation of the detector signal at the fourth harmonic (4Ω) of the tip oscillation frequency. The s-SNOM amplitude approach curves (fig. 7A) are acquired by measuring the demodulated detector signals as a function of the tip-sample distance at higher harmonic of the tip tapping frequency. This signal is maximum at small tip-sample distances and decreases rapidly as the tip is pulled away from the sample due to effective signal demodulation, which guarantees precise alignment of the excitation laser to the tip. Near-field approach curves (fig. S7A) were recorded routinely before either TEPL/TERS spectra or near-field images are taken.

For TERS/TEPL measurements, the $MoS_2$-$WS_2$ lateral heterostructure sample is excited by the laser and the forward scattered emission is collected by the parabolic mirror of the s-SNOM and dispersed using a 328 mm focal length Andor spectrometer and imaged with liquid nitrogen cooled silicon EMCCD camera (Andor iXon). Tip-sample distance dependence of TEPL/TERS signal of the ML heterostructure $MoS_2$-$WS_2$ is always checked before any measurement (fig. S7B). To obtain normalized TEPL spectra, the tip is positioned on a reference area on the substrate far away from the desired sample flake to record a reference spectrum. Then a spectrum is taken on the sample surface. A normalized spectrum is achieved by subtracting the reference spectra from the spectra taken on the sample. For hyperspectral TEPL/TERS nanoimaging we record normalized spectra at each pixel (x, y) of a 2D area of the sample surface (shown in Fig. 1). The capability to provide spatial and spectral information simultaneously with spatial resolution limited only by the apex radius of the probe tip (typically ~20 nm) marks the extraordinary advantage of TEPL/TERS hyperspectral nanoimaging. An additional advantage of this hybrid s-SNOM and TEPL/TERS nano-imaging setup is the ability to easily and routinely align the excitation beam guided by near-field approach curves.

**AC-HRSTEM:** As-grown $MoS_2$-$WS_2$ heterostructures and $Mo_xW_{1-x}S_2$ alloys was transferred either onto Quantifoil gold grid or *in situ* heating chip (E-AHBC) for Aduro holder (Protochips), using conventional PMMA-based wet chemical transfer (for Quantifoil grid) and PMMA-based deterministic transfer (for *in situ* heating chip). (*46*) For imaging, FEI Titan[3] G2 S/TEM 60-300 operated at 80 kV and high angle annular dark field (HAADF) detector was used. Lower voltage was used at high resolution to decrease irradiation damage. To enhance visibility and reduce noise in STEM images, all acquired high-resolution images were processed by Gaussian Blur filter (radius = 0.03 nm) using ImageJ software.

**Acknowledgments:**
A.F. and Y.A. thank Stefan Mastel and Andreas Huber (Neaspec co.) and Michael Serge (Andor co.) for many discussions and assistance in the initial development of the TERS/TEPL setup. Funding: A.F., S. K., Y. G. and Y.A. acknowledge support from Air Force Office of Scientific Research (AFOSR) Grant FA9559-16-1-0172. And the National Science Foundation (NSF) CAREER (for Y.A.) Grant No. 1553251. T.Z., E.K., K.F. and M.T. acknowledge support from the AFOSR through grant No. FA9550-18-1-0072 and the NSF-IUCRC Center for Atomically Thin Multifunctional Coatings (ATOMIC). W.S. and L.Y. are supported by the NSF CAREER Grant No. DMR-1455346 and the AFOSR Grant No. FA9550-17-1-0304. The computational resources have been provided by the Stampede of TeraGrid at the Texas Advanced Computing Center (TACC) through XSEDE

**Author contributions:** Y.A. and M.T. conceived and guided the experiments. A.F., S.K. and Y. A. carried out TEPL, TERS, and s-SNOM measurements. J.P.T and A.F. analyzed experimental data. Y.G. performed low temperature and far-field PL experiments. T.Z., E.K. and K.F. grew the samples and performed far-field PL/Raman and *AC-HRSTEM measurements*. K. F. performed temperature dependent HRSTEM experiments. W.S. and L.Y. performed *ab initio* MBPT calculations. All authors contributed to writing the manuscript.

**Competing interests:** The authors declare no competing interests.

**Supplementary Materials:**

Fig. S1. TEPL image of $MoS_2$ when the $WS_2$ peak is subtracted.
Fig. S2. High resolution hyperspectral image.
Fig. S3. Tip Enhanced Raman shift extracted at O-W-O stretching modes .
Fig. S4. PL peak energies over time indicate blueshift and broadening of PL emission after degradation.
Fig. S5. Strain and PL emission over time show less degradation at intersections.
Fig. S6. All 3 excitonic species ($X^0$, $X^-$, and XX) can be tracked to room temperature at both selected points ($WS_2$ edge and intersection) during the photo-oxidation process.
Fig. S7. Near-field signal and TEPL intensity dependence to tip-sample distance.

Table S1. PL intensities and intensity ratios of the various species reveal remarkably higher concentrations of trions and biexcitons on Day 1 at the corner potential well region (Point 1) compared to the $WS_2$ edge region (Point 2) in addition to significantly higher persistence over time of these species at Point 1.

## Figures

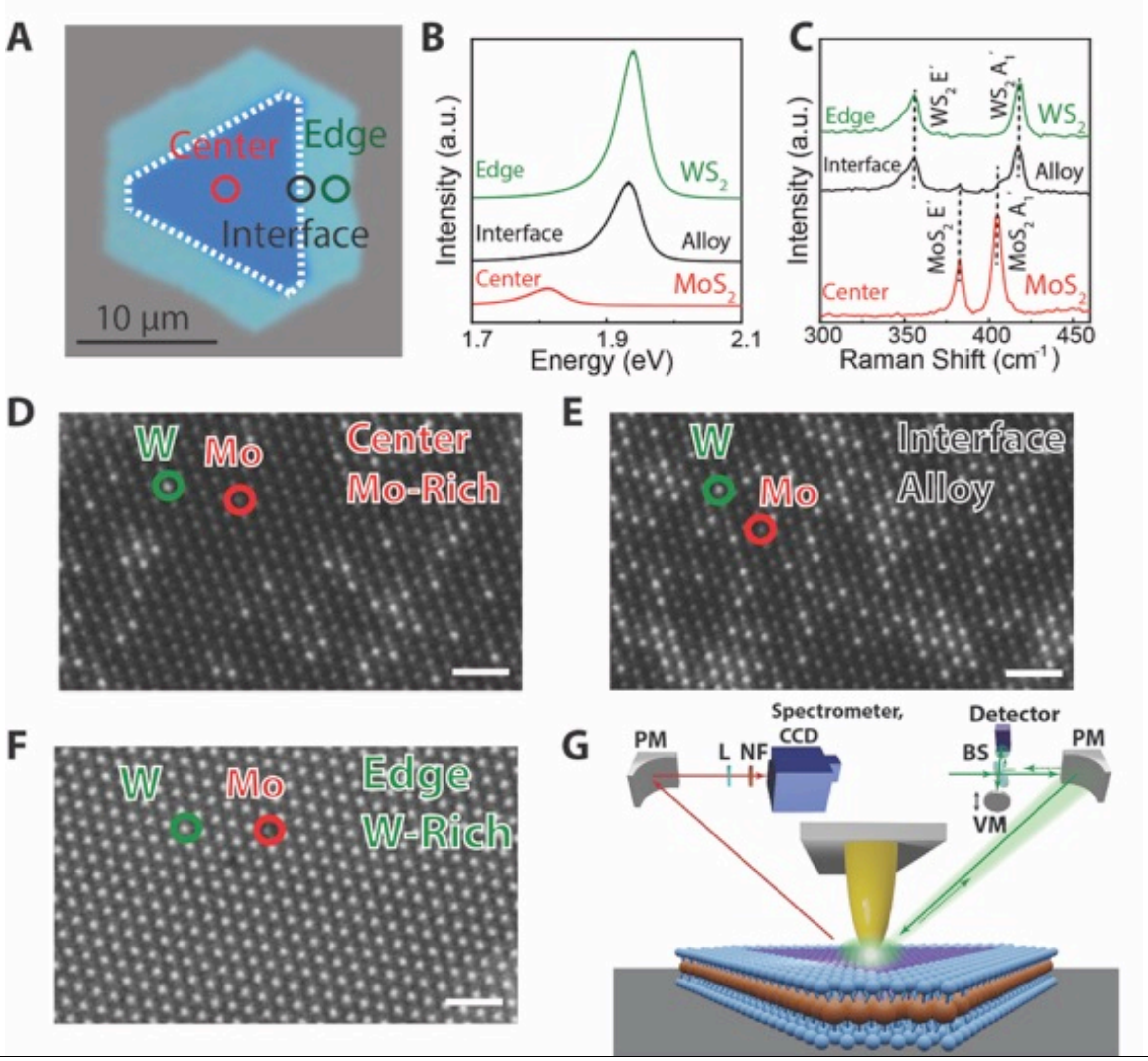

**Fig. 1. Monolayer $MoS_2$-$WS_2$ lateral heterostructures**. (**A**) An optical image of as-grown in-plane $MoS_2$-$WS_2$ heterostructures on $SiO_2$/Si substrate. Center, interface and edge regions of the flake are marked with red, black and green circles, respectively. The interface region is highlighted in white broken lines. (**B**) Photoluminescence (PL) spectra of the center, interface and edge regions of the $MoS_2$-$WS_2$ lateral heterostructures. (**C**) Raman spectra of the center, interface and edge regions of the $MoS_2$-$WS_2$ lateral heterostructures. (**D-F**) Typical AC-HRSTEM images acquired from the center, interface, and edge areas of the $MoS_2$-$WS_2$ lateral heterostructures. Atoms with brighter and darker contrast are W (green circle) and Mo (red circle), respectively. The STEM-high-angle annular dark-field (HAADF) characterization confirms the center region is Mo-rich, the edge region is W-rich, and the interface displays an intermixing of Mo and W, forming an alloyed $Mo_xW_{1-x}S_2$ composition transition region. Scale bar is 1 nm. (**G**) Schematics of integrated s-SNOM/TEPL/TERS experimental setup (PM=parabolic mirror, NF=neutral density filter, BS=beam splitter, VM=vibrating mirror).

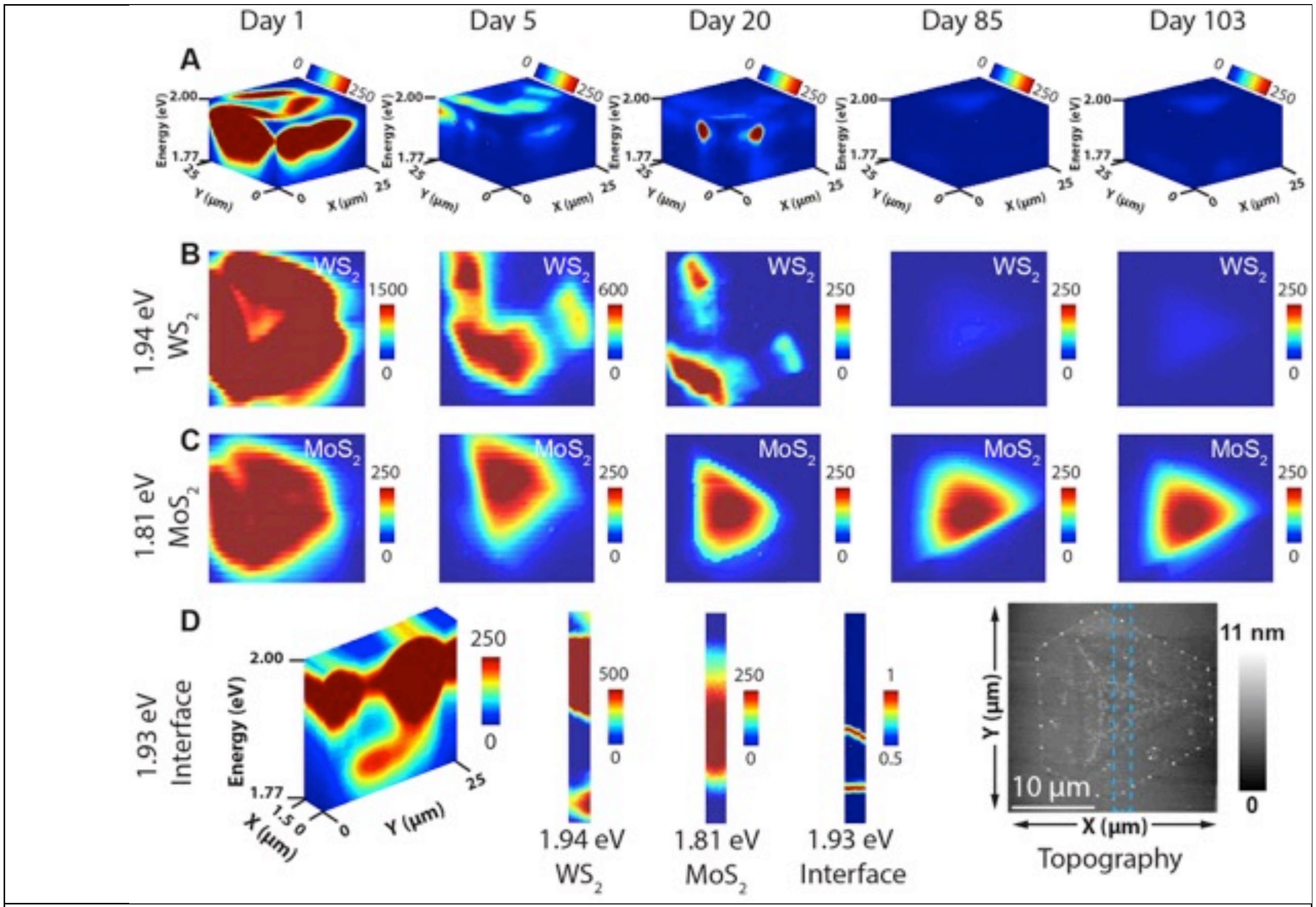

**Fig. 2. Hyperspectral spatial TEPL nanoimaging showing time evolution of exciton emission in heterostructure monolayer MoS2/WS2.** (**A**) TEPL hyperspectral 3D data cube taken as a function of time (days 1-103). (**B**) cross-section cut of TEPL map taken at 1.94 eV and (**C**) 1.81 eV corresponding to the exciton peaks expected for WS2 and MoS2 respectively. (**D**) High-resolution TEPL hyperspectral 3D data cube of flake taken in blue dashed area shown in topography and cross-section cut of high resolution TEPL map taken at 1.94 eV, 1.81 eV, and 1.93eV corresponding to the exciton peaks expected for WS2 and MoS2 and interface respectively.

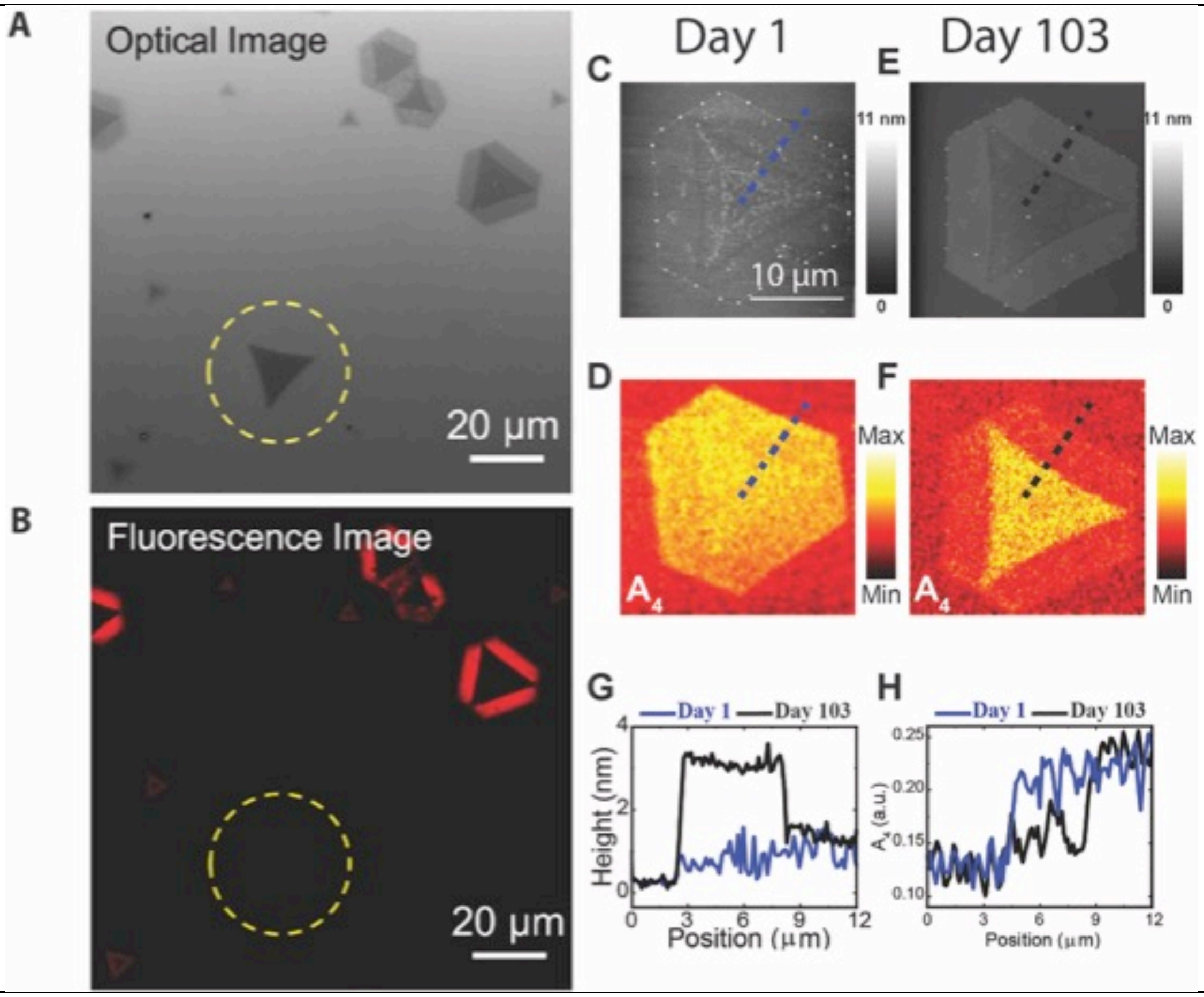

**Fig. 3.** (**A**) Optical image of MoS2-WS2 lateral heterostructures acquired on day 103. The laser-irradiated flake is distinguished and marked with a yellow dashed circle, which exhibit a very different optical contrast compared with other unexposed flakes. (**B**) The corresponding fluorescence image acquired using a band-pass filter centered at 630 nm (also acquired on the same day). The irradiated flake, fluorescence emission from edge region is completely quenched. Topography and 4th harmonics near-field images and line profiles of the MoS2-WS2 monolayer heterostructure taken at day 1 (**C-D**) and day 103 (**E-F**). (**G-H**) topography and fourth harmonic near-field amplitude change of MoS2-WS2 monolayer heterostructure between day 1 and day 103.

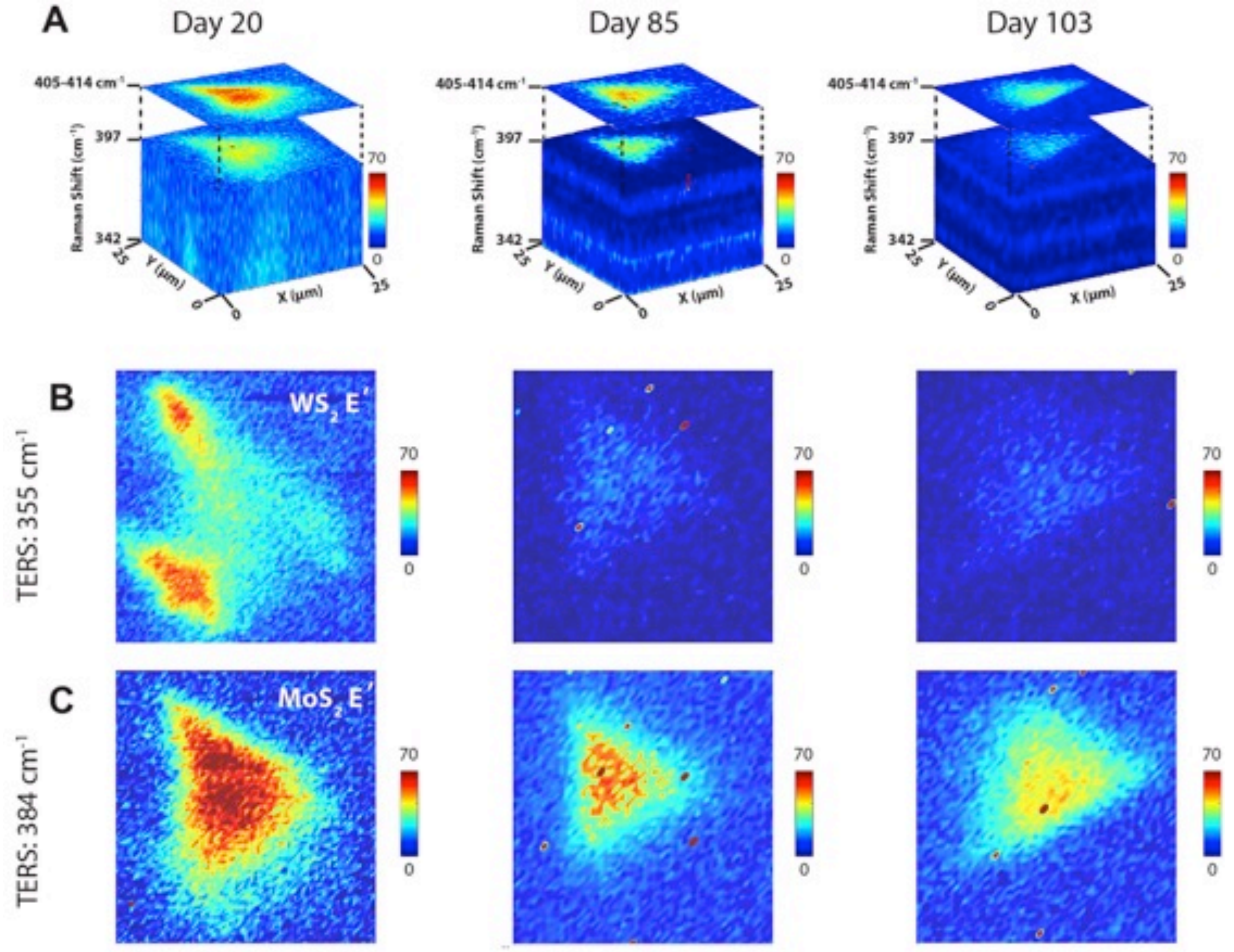

**Fig. 4. Spatial evolution of Raman shift as a function of time in photooxidized heterostructure monolayer $MoS_2/WS_2$.** (**A**) Hyperspectral TERS 3D data cube shown at selected days. (**B**) cross-section cut of TERS map taken at 355 cm$^{-1}$(**C**) and 384 cm$^{-1}$ corresponding to the Raman shifts expected for $WS_2$ and $MoS_2$ respectively on days 20, 85 and 103.

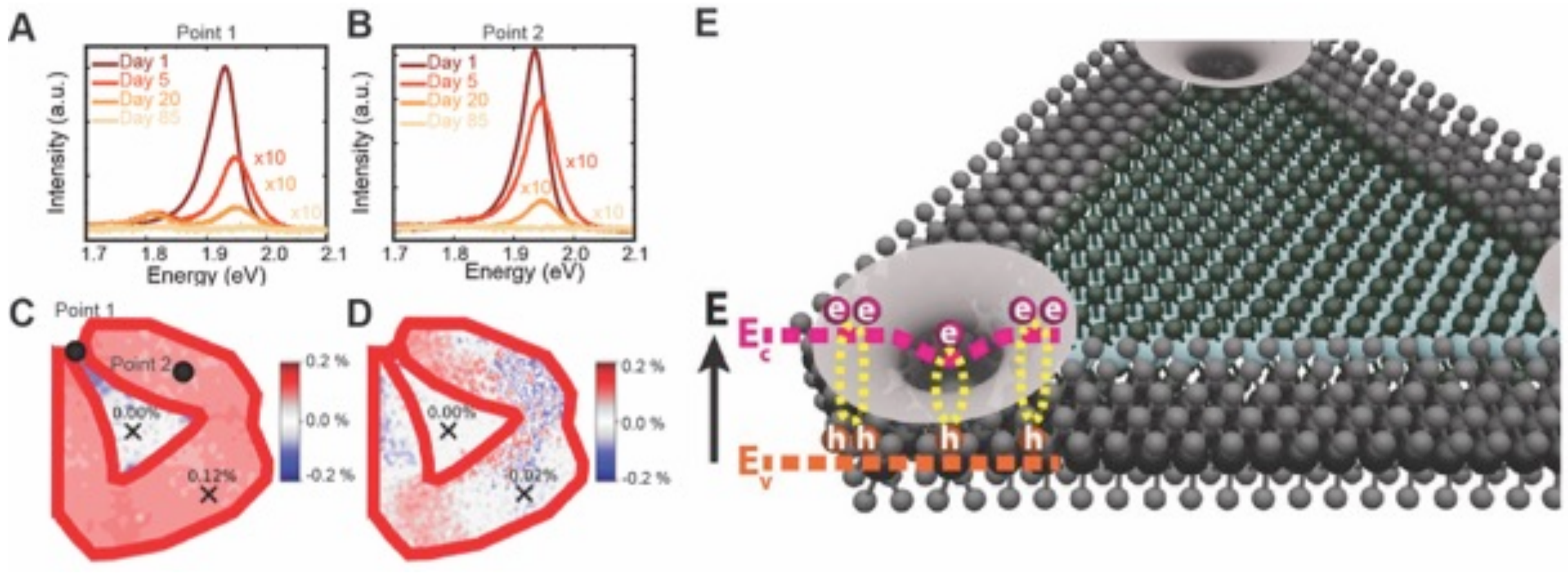

Fig 5. (A-B) Spectral shift with photooxidation at points 1&2 shown in black dots in (C) as a function of time on the $WS_2/MoS_2$ ML heterostructure. (C) and (D) Spatial strain distribution by fitting energy of local PL emission peaks to *ab initio* MBPT results corresponding to heterostructure TEPL images on day 1 and day 20, respectively. (E) Schematic plot of the accumulating effect of excitons by strain around a corner of samples.

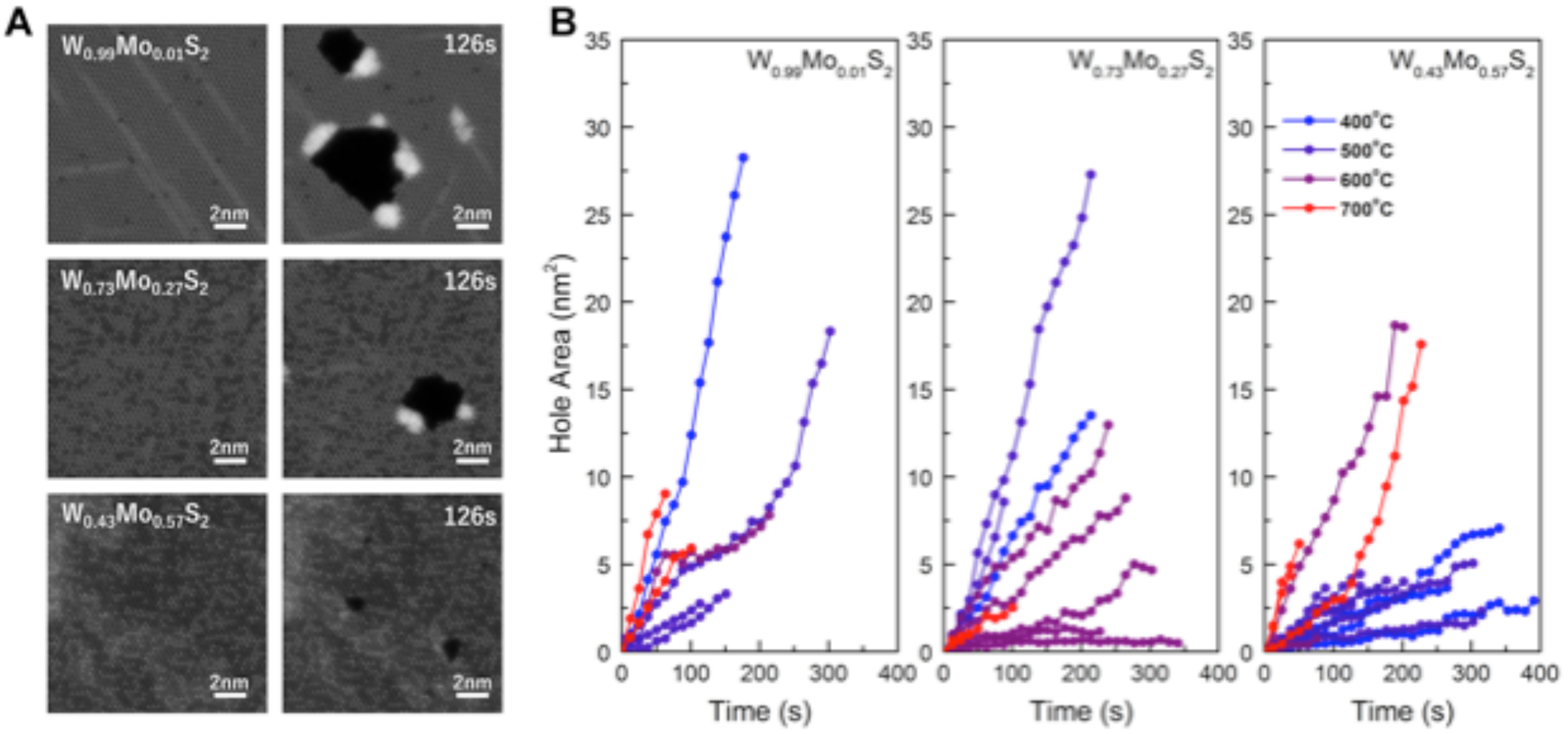

**Fig. 6. *In situ* HRSTEM stability analysis of $Mo_xW_{1-x}S_2$ alloys at elevated temperature.** (**A**) HRSTEM images of $Mo_xW_{1-x}S_2$ before and after 126s of continuous scanning at 400 °C. (**B**) Time-dependent hole area evolution plot at different test temperatures. The sample with the highest alloying degree ($W_{0.43}Mo_{0.57}S_2$) exhibits the highest thermal stability.